\documentclass[aps,prb,showpacs,preprint,superscriptaddress,amsfonts,amsmath,amssymb]{revtex4-1}

\usepackage{graphicx}
\usepackage{dcolumn}
\usepackage{bm}
\usepackage{accents}
\usepackage{xcolor}
\usepackage{appendix}
\begin{document}


\title{Quasiparticle Band Gaps, Excitonic Effects, and Anisotropic Optical Properties of Monolayer Distorted 1-T Diamond-Chain Structure
ReS$_2$ and ReSe$_2$}

\author{Hong-Xia Zhong}
\affiliation{Department of Physics, Washington University in St.
Louis, St. Louis, Missouri, 63130, USA}

\affiliation{State Key Laboratory for Mesoscopic Physics and
Department of Physics, Peking University, Beijing 100871, China}

\author{Shiyuan Gao}
\affiliation{Department of Physics, Washington University in St.
Louis, St. Louis, Missouri, 63130, USA}

\author{Jun-Jie Shi}

\affiliation{State Key Laboratory for Mesoscopic Physics and
Department of Physics, Peking University, Beijing 100871, China}

\author{Li Yang}
\affiliation{Department of Physics, Washington University in St.
Louis, St. Louis, Missouri, 63130, USA}

\begin{abstract}
We report many-body perturbation theory calculations of
excited-state properties of distorted 1-T diamond-chain monolayer
rhenium disulfide (ReS$_2$) and diselenide (ReSe$_2$). Electronic
self-energy substantially enhances their quasiparticle band gaps
and, surprisingly, converts monolayer ReSe$_2$ to a direct-gap
semiconductor, which was, however, regarded to be an indirect one
by density-functional-theory calculations. Their optical
absorption spectra are dictated by strongly bound excitons. Unlike
hexagonal structures, the lowest-energy bright exciton of
distorted 1-T ReS$_2$ exhibits a perfect figure-8 shape
polarization dependence but those of ReSe$_2$ only exhibit a
partial polarization dependence, which results from two
nearly-degenerated bright excitons whose polarization preferences
are not aligned. Our first-principles calculations are in
excellent agreement with experiments and pave the way for
optoelectronic applications.
\end{abstract}


\maketitle


\section{Introduction}

To overcome the zero band gap of graphene
\cite{neto2009electronic}, two-dimensional (2D) transition-metal
dichalcogenides (TMDCs) semiconductors have attracted significant
attention \cite{wang2012electronics, mak2010atomically, wang2010,
yao2012, feng2012}. Thanks to the very thin atomic thickness (a
few {\AA}), pristine interfaces without out-of-plane dangling
bonds, and considerable band gaps (1-2 eV) \cite{mos2-1, mos2-2,
wang2012electronics, qiu2013optical, yang2013}, TMDCs are
potentially beneficial for eliminating short channel effects
\cite{liu2012channel}, lowering interface state densities,
reducing surface roughness scattering \cite{geim2013van}, and
low-power digital applications \cite{radisavljevic2011single,
chuang2014high, liu2013role, das2012high}. However, because of
interlayer interactions and variations of screening, the
electronic structures of many hexagonal TMDCs undergo substantial
variations with different stacking layer numbers. For example, a
crossover from direct band gaps in monolayers to indirect band
gaps in multilayers has been observed in MoS$_2$
\cite{mak2010atomically, wang2010, zhang2014direct}, limiting the
applicability of TMDCs in optoelectronic devices.

Layered crystals of rhenium disulfide (ReS$_2$) and diselenide
(ReSe$_2$) are a new family of 2D TMDCs semiconductors, which have
been successfully fabricated recently \cite{tongay2014monolayer,
yang2014layer, wolverson2014raman}. Unlike hexagonal TMDCs,
ReX$_2$ (X= S, Se) crystallizes in a distorted 1-T diamond-chain
structure with the triclinic symmetry, as a result of charge
decoupling from an extra valence electron of Re atoms
\cite{tongay2014monolayer, yang2014layer, wolverson2014raman}.
This structural distortion leads to a much weaker interlayer
coupling. Consequently, the band renormalization is absent and
bulk ReX$_2$ behaves as electronically and vibrationally decoupled
monolayers \cite{tongay2014monolayer, yang2014layer}. Such a
vanishing interlayer coupling in ReX$_2$ structures enables
probing 2D-like systems without the need of monolayer or few
layers, overcoming the challenge of preparing large-area,
single-crystal monolayers \cite{tongay2014monolayer}. Furthermore,
the structure distortion of 1T-ReX$_2$ makes it exhibit unique
anisotropic optical, electrical, and mechanical properties
\cite{tongay2014monolayer, wolverson2014raman}, with considerable
interests for various applications in polarization controller,
liquid crystal displays, 3D visualization techniques,
(bio)dermatology, and in optical quantum computers
\cite{nan2009linear,knill2001scheme}.

Despite these unique properties and potential applications of
distorted 1-T diamond-chain ReX$_2$ structures, we have very
limited knowledge about their fundamental excited-state
properties, such as quasiparticle band gaps, optical spectra, and
excitonic effects. In particular, currently available experimental
measurements are diverse due to sample qualities and unavoidable
environment effects. In this sense, a reliable and parameter-free
calculation to accurately capture essential electron-electron
(\emph{e-e}) and electron-hole (\emph{e-h}) interactions in
excited-state properties is highly desirable for understanding
fundamental physics and motivating applications of few-layer
ReX$_2$.

In this work, we employ the first-principles GW-Bethe-Salpeter
Equation (BSE) approach to study two typical distorted 1-T
diamond-chain TMDCs, i.e., monolayer ReS$_2$ and ReSe$_2$. We
observe significant many-electron effects that dictate their
excited-state properties. Self-energy corrections not only enlarge
band gaps but also convert suspended monolayer ReSe$_2$ to be a
direct-gap semiconductor. Excitonic effects are prominent and the
BSE-calculated absorption spectra are in excellent agreement with
measurements. In particular, because of the reduced triclinic
symmetry, monolayer ReS$_2$ and ReSe$_2$ possess anisotropic
optical responses within the near-infrared frequency regime. These
excited-state properties within the interesting frequency regime
may be useful for optoelectronic applications.

The remainder of this paper is organized as follows: In Sec. II,
we introduce the atomic structures of monolayer ReS$_2$ and
ReSe$_2$, and our computational approaches. In Sec. III, the
quasiparticle band energies, band gaps, and the transition from
indirect to a direct band gap, are presented. In Sec. IV, we
discuss the optical absorption spectra and excitonic effects. In
Sec. V, we focus on the anisotropic optical responses of excitons.
In Sec. VI, we summarize our studies and conclusion.

\section{Atomic Structure and Computational Methods}
\begin{figure}
\includegraphics[width= 10.0cm]{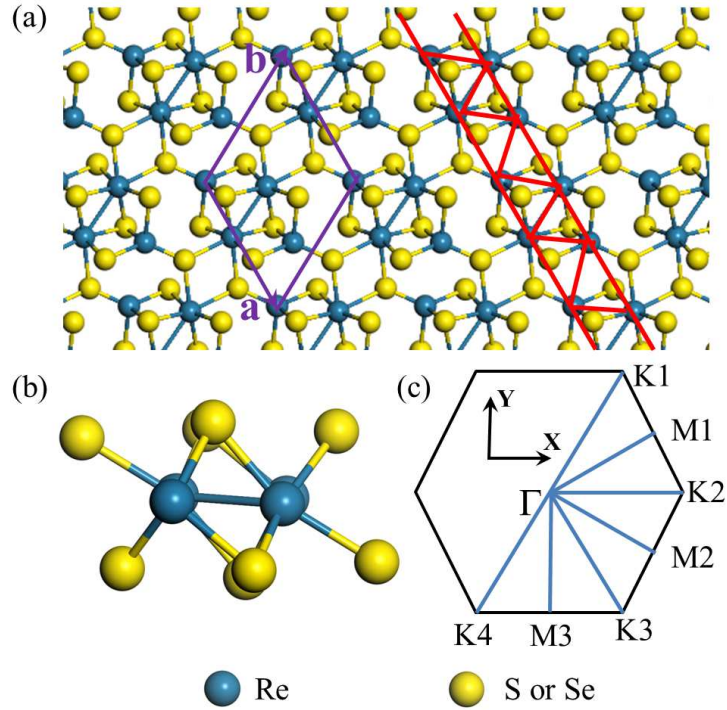}
\caption{(a) Top view of atomic structure of distorted 1-T
diamond-chain monolayer ReX$_2$. The clustering of Re atoms forms
the zigzag chains along lattice vector \textbf{a} direction, as
shown in the red-line box. (b) The side view of monolayer ReX$_2$.
(c) The first Brillouin zone of monolayer ReX$_2$.} \label{fig:1}
\end{figure}

The ball-stick structures of monolayer distorted 1-T diamond-chain
ReS$_2$ and ReSe$_2$ are presented in Fig.~\ref{fig:1}. The unit
cell and unit vectors are marked by purple-colored vectors. Each
Re atom is clustering of ``diamond chains" composed of the zigzag
shape along the \textbf{a} direction as shown by red-colored lines
in Fig.~\ref{fig:1}(a)
\cite{tongay2014monolayer,horzum2014formation,
corbet2014field,yang2015tuning}. The side view of monolayer
ReS$_2$ or ReSe$_2$ is presented in Fig.~\ref{fig:1}(b). The
special feature is that, unlike hexagonal TMDCs, the S or Se atoms
of distorted 1-T diamond-chain structures are not all in the same
plane, substantially lowering the structure symmetry. The first
Brillouin zone (BZ) is plotted in Fig.~\ref{fig:1}(c), which is a
hexagon but with unequal side lengths as a result of the distorted
atomic structure. Consequently, the K1, K2, and K3 points are no
longer equivalent.

We fully relax the structures according to the force and stress
calculated by DFT with the Perdew, Burke, and Ernzerhof (PBE)
functional \cite{perdew1996generalized}, using the \textsc{Quantum
Espresso} package \cite{giannozzi2009quantum}. The ground-state
wave functions and eigenvalues are obtained from the DFT/PBE with
norm-conserving pseudopotentials \cite{troullier1991efficient}.
The plane-wave basis is set with a cutoff energy of 60 Ry with a
$8\times8\times1$ $k$-point grid. A vacuum space between
neighboring layers is set to be more than 25 {\AA} to avoid the
interactions between layers. Based on these parameters, our
calculated lattice constant \textbf{a} is 6.43 (6.78) {\AA} and
the lattice constant \textbf{b} is 6.53 (6.66) {\AA} for ReS$_2$
and ReSe$_2$, respectively. They are consistent with previous
calculations \cite{tongay2014monolayer,wolverson2014raman}.

The quasiparticle energies and band gaps are calculated by the GW
approximation within the general plasmon pole (GPP) model
\cite{hybertsen1986electron}. The involved unoccupied conduction
band number for calculating the dielectric function and
self-energy is about ten times of the occupied valence band
number. In solving the BSE, we use a finer $k$-point grid of
$32\times32\times1$ for converged excitonic states
\cite{rohlfing2000electron}. All the GW-BSE calculations are
performed with the BerkeleyGW code \cite{berkeleyGW} including the
slab coulomb truncation scheme to mimic suspended monolayer
structures \cite{ismail2006truncation, rozzi2006exact}. For
optical absorption spectra, only the incident light polarized
parallel with the plane is considered due to the depolarization
effect \cite{spataru2004quasiparticle, yang2007enhanced}.

\section{Quasiparticle Energies and Band Gap}

The DFT calculated band structures of monolayer ReS$_2$ and
ReSe$_2$ are presented in Fig.~\ref{fig:2}. That with spin-orbit
coupling (SOC) included is presented in the appendix. As shown in
Fig.~\ref{fig:2}(a), monolayer ReS$_2$ exhibits a 1.36 eV direct
band gap at the $\Gamma$ point. This is consistent with a recent
work \cite{tongay2014monolayer} reporting a value of 1.43 eV. On
the other hand, at the DFT level, monolayer ReSe$_2$ is observed
to be an indirect-gap semiconductor; the conduction band minimum
(CBM) is located at the $\Gamma$ point while the valence band
maximum (VBM) is slightly away from the $\Gamma$ point, as shown
in Fig.~\ref{fig:2}(b) and its inset. The DFT-calculated indirect
band gap of monolayer ReSe$_2$ is about 1.22 eV and the direct gap
at the $\Gamma$ point is slightly larger, about 1.25 eV. These are
also consistent with previous DFT calculations
\cite{yang2014layer,wolverson2014raman, yang2015tuning}.

However, it is well known that DFT usually underestimates the band
gap of semiconductors. Therefore, we have performed the GW
calculation to obtain reliable quasiparticle band gaps of
monolayer ReS$_2$ and ReSe$_2$. Similar to those found in other
monolayer 2D semiconductors, such as monolayer hexagonal TMDCs
(MoS$_2$) \cite{mos2-1, mos2-2, qiu2013optical,yang2013} and
phosphorene \cite{tran2014layer}, significant self-energy
enhancements are observed in ReX$_2$; at the ''single-shot"
G$_0$W$_0$ level, the quasiparticle band gap of monolayer ReS$_2$
is increased to be 2.38 eV and that of monolayer ReSe$_2$ is
increased to 2.09 eV, as listed in Table I. These enhanced
many-electron effects are from the depressed screening and reduced
dimensionality of suspended 2D semiconductors
\cite{qiu2013optical, tran2014layer, mos2-1,
shi2013quasiparticle}.

Interestingly, the self-energy not only enlarges the band gap
value but also may modify the band topology of ReX$_2$. This is
particularly significant for monolayer ReSe$_2$, which is an
indirect-gap semiconductor at the DFT level, as shown in
Fig.~\ref{fig:2}(b). However, \emph{e-e} interactions lower the
valence band energy at the $\Lambda$ point by about 50 meV more,
compared to that of the $\Gamma$ point. Therefore, the VBM is
shifted to the $\Gamma$ point, making monolayer ReSe$_2$ a
direct-gap semiconductor, as listed in the blue-colored VBM in
table II. However, it has to be pointed out that the indirect to
direct transition may not be very sharp because the energy
difference at the $\Gamma$/ $\Lambda$ point are very small and
external perturbations may change the conclusion. On the other
hand, it shall be true that the many-electron self-energy will
increase the valence band energy at the $\Lambda$ point and, at
least, make the top of valence band almost degenerated. This will
affect photoluminescence (PL) experiments.

These different self-energy corrections are resulted from the
nature of the involved electronic states at the $\Gamma$ and
$\Lambda$ points. For example, a major contribution to the
self-energy is from the screened-exchange interaction
($\Sigma_{SEX}$) \cite{hybertsen1986electron}, which is strongly
affected by the spatial localization of electronic states. This
can be seen from its static form \cite{hybertsen1986electron}:
\begin{equation}
\Sigma_{SEX}(r,r')=-\Sigma_{nk}^{occ}\phi_{nk}(r)\phi^*_{nk}(r')W(r,r'),
\end{equation}
where the subscripts $n$ is the band index and $k$ is the sampling
points in the reciprocal space, $\phi_{nk}$ is the electronic wave
function, and $W$ is the screened Coulomb interaction. Obviously,
a smaller distance between $r$ and $r'$ enhances the overlap
between wave functions and screened Coulomb interactions, giving
rise to a larger self-energy. This non-uniform self-energy
correction is similar to what have been observed in hexagonal
TMDCs, i.e., the different self-energy corrections at the K point
and the zone center of monolayer MoS$_2$ \cite{qiu2013optical}.
Following this idea, we have checked the projected density of
states (PDOS) of electronic states at the $\Gamma$ and $\Lambda$
points of ReSe$_2$, as listed in Table~\ref{table2}. The
$\Lambda_v$ valence state has more localized $d$-electron
component (86\%) than that the $\Gamma_v$ valence state (78\%). As
a result, quasiparticle energy of the $\Lambda_v$ state is
enhanced more, meaning its energy level is lowered more and
resulting in a switch of the VBM.

This direct quasiparticle band gap is not conflicted with
experimental measurements, in which rather weak PL peak intensity
is observed in monolayer ReSe$_2$ and the PL peak of ReSe$_2$
increases monotonically with increasing the layer number
\cite{zhao2015interlayer}. This measurement seems to hint an
indirect-gap of ReSe$_2$. However, different from our calculated
suspended and neutral cases, these measured samples are on
substrates and are inevitably doped. Thus their self-energy
corrections will be substantially reduced under these conditions,
as shown in WSe$_2$ and MoS$_2$ \cite{louie2014,yang2015}, meaning
their measured samples may have a slightly indirect band gap. It
has to be pointed out that this small energy difference is within
the intrinsic error bars of DFT and GW methods. Thus more accurate
experiments of suspended samples are important for conclusive
results.

Finally, previous works have shown that the self-consistent GW
(sc-GW) scheme beyond single-shot calculations may be necessary
for 2D semiconductors \cite{shi2013quasiparticle, qiu2013optical,
tran2014layer}. Therefore, we perform one self-consistent update
to the Green's function G; the quasiparticle band gap is further
increased to 2.82 eV for monolayer ReS$_2$ and 2.45 eV for
monolayer ReSe$_2$, as listed in Table~\ref{table1}. Similar to
monolayer MoS$_2$ and black phosphorus, we find that more
self-consistent steps only slightly change the band gap ($\sim 0.1
eV$) and we stop at the sc-G$_1$W$_0$ level. Finally, the SOC is
significant in ReX$_2$ structures (around 150 meV of lowering the
band gap) \cite{tongay2014monolayer}. Considering all these
factors, we compute the quasiparticle band gaps and summarize them
in the last column of Table~\ref{table1}. In the following, all
our discussions and the BSE calculations are based on the
finalized sc-G$_1$W$_0$ results with SOC included.

\section{Exciton Effects on Optical Absorption Spectra}

\begin{figure}
\includegraphics[width=12.0cm]{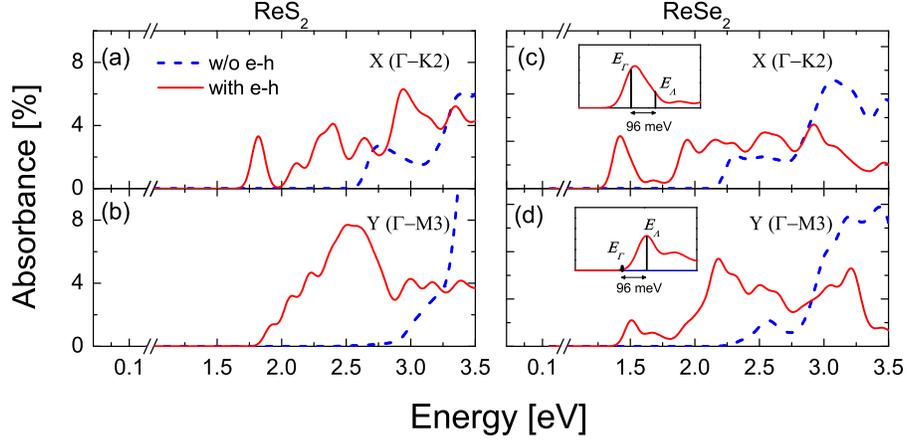}
\caption{Optical absorption spectra of monolayer ReS$_2$ (left)
and ReSe$_2$ (right) for the incident light polarized along the
$\Gamma$-K$_2$ direction ((a) and (c)) and the $\Gamma$-M$_3$
direction ((b) and (d)). The inset in (c) and (d) zoom in the
lowest-energy absorption peaks, which is composed of two
adjacent excitons ($E_\Gamma$ and $E_\Lambda$) with an energy
difference 96 meV in monolayer ReSe$_2$. We employ 50 meV Gaussian
smearing in these plots.} \label{fig:3}
\end{figure}

Because optical spectra of ReX$_2$ are anisotropic, we first focus
on those with the incident light polarized along the $\Gamma$-K2
(x) directions because the corresponding optical absorption
spectra can exhibit most features. Those of other polarizations
will be discussed in Section V. First, the single-particle optical
absorption spectra for monolayer ReS$_2$ and ReSe$_2$ are
presented in Figs.~\ref{fig:3}(a) and (c) (blue dashed lines). As
expected, the main optical absorption shoulder starts from the
quasiparticle band gap, which are around 2.7 eV and 2.3 eV, for
ReS$_2$ and ReSe$_2$, respectively.

Since \emph{e-h} interactions are known to be crucial for
obtaining reliable optical spectra in 2D semiconductors, we have
solved the BSE to include them and present the optical absorption
spectra in Fig.~\ref{fig:3} by red solid lines. Similar to other
2D semiconductors, all main optical features are dominated by
excitonic states in monolayer ReX$_2$. For example, after
including \emph{e-h} interactions, the first absorption peak of
ReS$_2$ is located at 1.63 eV as shown in Fig.~\ref{fig:3}(a),
which is a strongly bound exciton with a 1.07-eV \emph{e-h}
binding energy. Monolayer ReSe$_2$ exhibits similarly enhanced
excitonic effects as shown in Fig.~\ref{fig:3}(c) with the
absorption peak at around 1.42 eV and exciton binding energy of
0.87 eV.

\begin{figure}
\includegraphics[width=11.0cm]{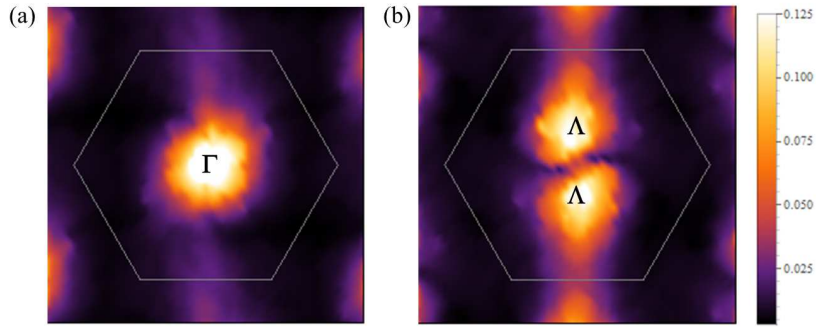}
\caption{The exciton amplitude distribution of the exciton
$E_\Gamma$ (a) and that of the $E_\Lambda$ (b) (see
Fig.~\ref{fig:3}) in monolayer ReSe$_2$ in reciprocal space.}
\label{fig:4}
\end{figure}

It has to be pointed out that the lowest-energy peak (1.42 eV) in
the optical absorption spectrum of monolayer ReSe$_2$ is actually
composed of two bright excitons ($E_\Gamma$ and $E_\Lambda$), as
elucidated in the insert of Fig.~\ref{fig:3}(c). The energy
spacing of these two excitons is only about 96 meV. These two
nearly degenerated excitons are resulted from the indirect-gap
nature of ReSe$_2$. To see this, we have plotted the distributions
of these exciton amplitudes in the reciprocal space (see details
in Ref.~\onlinecite{qiu2013optical}). As shown in
Fig.~\ref{fig:4}, the exciton $E_\Gamma$ is mainly from the
transitions around the zone center (the $\Gamma$ point), while the
exciton $E_\Lambda$ is mainly from those around the $\Lambda$
points. Since the energies of maxima of the valence band at the
$\Gamma$ and $\Lambda$ points are very close to each other, the
formed excitons obviously have very similar energies as well.

Recently, several PL measurements of monolayer ReS$_2$ and
ReSe$_2$ have been performed. The measured broad PL peak position
is about 1.50 to 1.65 eV at room temperature for ReS$_2$
\cite{tongay2014monolayer, horzum2014formation} and 1.47 eV for
ReSe$_2$ \cite{yang2015tuning}. These experimental observations
are in excellent agreement with our first-principles calculations,
which we calculate as 1.6 eV and 1.4 eV for ReS$_2$ and ReSe$_2$,
respectively. As shown in many previous works \cite{rubio2006,
komsa2012, louie2014}, the quasiparticle self-energy and
\emph{e-h} interactions partially cancel each other, making the
final ``optical" gap not much larger than the DFT-calculated band
gap.

Finally, we have to address that Coulomb truncation is crucial for
all above calculations \cite{komsa2012, wirtz2013}. Without
Coulomb truncation, both self-energy corrections and excitonic
effects will be reduced because the actual simulation object is
periodic structures with a large interlayer distance. For
instance, for monolayer black phosphorus \cite{tran2014layer}, the
GW calculation with Coulomb truncation gives the band gap as 2 eV
while that without truncation gives the value of 1.7 eV.

\section{Anisotropic Exciton}

Unlike hexagonal TMDCs, monolayer ReS$_2$ and ReSe$_2$ are the
distorted 1-T diamond-chain with the triclinic symmetry. As a result, the
optical absorption may be also anisotropic. This has been
previously examined at the DFT level \cite{ho2004optical}, where
it is determined that the absorption spectra are anisotropic with
respect to the polarization direction of the incident light.
However, due to the incorrect band gaps and the short of
\emph{e-h} interactions, those DFT-calculations cannot be directly
compared with experiments for providing reliable understanding.
This motivates us to study the anisotropic optical responses at
the level of including quasiparticle energy and excitonic effects.

Let us first focus on the case of monolayer ReS$_2$. Its
symmetry-induced anisotropic optical response can be seen even at
the single-particle optical absorption level. As shown in
Figs.~\ref{fig:3}(a) and (b), the single-particle optical
absorption spectrum begins at near 2.7 eV for the incident light
polarized along the $\Gamma$-K2 (x) direction, while it becomes
significant at 2.9 eV for the light polarized along the
$\Gamma$-M3 (y) direction. Previous studies have discussed this
optical anisotropy \cite{ho2004optical,ho2001plane} and it is
attributed to the transitions from non-bonding Re 5$d$ $t_{2g}$ to
5$d$ $t_{2g}^*$ and to anti-bonding chalcogen $p$ and $\sigma$
states \cite{ho2004optical}. In Table~\ref{table2}, we have listed
the PDOS of band-edge states of monolayer Res$_2$ and ReSe$_2$. It
is clear that both valence and conduction band edges are mainly
made of Re 5$d$ states and chalcogen $p$ states. For example, the
$\Gamma_v$ state of ReS$_2$ has 11 percents of $p$ state of S
atoms and 80 percents of $d$ state of Re atoms. This is consistent
with previous studies \cite{ccakir2014doping}.

The inclusion of \emph{e-h} interactions does not essentially
change the anisotropic optical response. As shown by red solid
lines in Figs.~\ref{fig:3}(a) and (b), excitonic effects only
lower the absorption edge but keep the anisotropy. For example,
the lowest-energy excitonic peak of monolayer ReS$_2$ is at 1.63
eV for the incident light polarized along the $\Gamma$-K2 (x)
direction and the absorption edge is located at 1.9 eV for the
incident light polarized along the $\Gamma$-M3 (y) direction. Thus
monolayer ReS$_2$ may work as an optical polarizer covering the
frequency range from 1.6 eV to 1.9 eV, which is very interesting
for near-infrared applications.

\begin{figure}
\includegraphics[width=10.0cm]{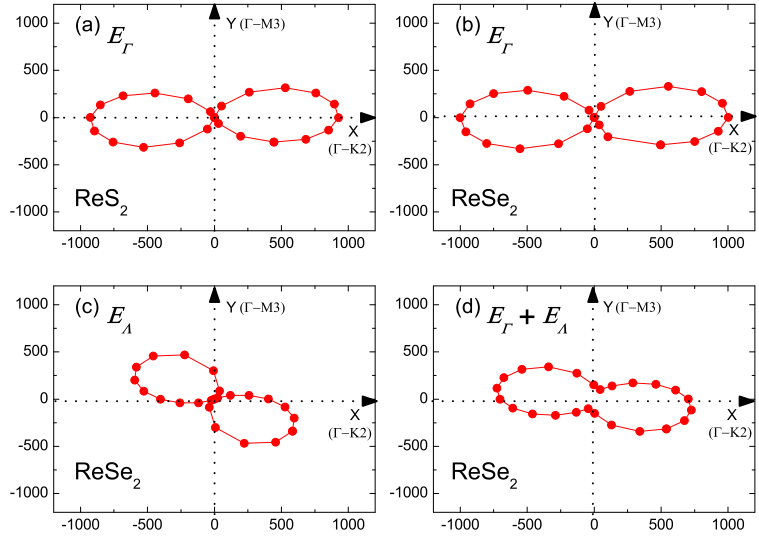}
\caption{Polarization-dependent oscillator strength (arbitrary
unit) of excitons in monolayer ReS$_2$ (a), the exciton $E_\Gamma$
(b) and $E_\Lambda$ (c) and their combined oscillator strength (d)
in monolayer ReSe$_2$ with a 50-meV smearing to mimic the
absorption spectrum.} \label{fig:5}
\end{figure}

In order to have a complete picture of anisotropies of these
important bright excitons, we present their optical oscillator
strength with respect to the polarization angle of the incident light
in Fig.~\ref{fig:5}. As expected, the optical oscillator strength
of the prominent exciton of monolayer ReS$_2$ exhibits a figure-8
shape (Fig.~\ref{fig:5}(a)), which represents the spatial
anisotropic optical response. For different directions, the
optical oscillator strength can differ by two orders of magnitude.
For instance, the intensity along $\Gamma$-K2 (x) direction is
about 800 times stronger than that along $\Gamma$-M3 (y)
direction. This anisotropic response can easily be observed by
both optical absorption and PL experiments.

Monolayer ReSe$_2$ exhibits a similar anisotropic optical
response, with some subtle differences. As shown in
Figs.~\ref{fig:3}(c) and (d), the optical absorption spectra of
monolayer ReSe$_2$ are different with respect to the polarization
direction of incident light. However, the lowest-energy absorption
peak (around 1.4 eV) along the $\Gamma$-K2 (x) direction does not
completely disappear for the incident light polarized along the
$\Gamma$-M3 (y) direction. This is different from the case of
ReS$_2$. This complication arises from the fact that the
lowest-energy peak of ReSe$_2$ actually consists of two excitons
($E_\Gamma$ and $E_\Lambda$), instead of one in ReS$_2$, as shown
in the insert of Figs.~\ref{fig:3}(c) and (d). $E_\Gamma$ behaves
similar to the lowest exciton in ReS$_2$, which is bright when the
incident light is polarized along the $\Gamma$-K2 (x) direction,
and dark along the $\Gamma$-M3 (y) direction. $E_\Lambda$, on the
other hand, is not completely dark along the $\Gamma$-M3 (y)
direction. Thus the lowest-energy peak does not have an extremely
high polariztion anisotropy.

This can be better understood by plotting the oscillator strength
of the excitons $E_\Gamma$ and $E_\Lambda$ seperately, which is
shown in Figs.~\ref{fig:5}(b) and (c). Interestingly, both
excitons, $E_\Gamma$ and $E_\Lambda$, exhibit significant spatial
anisotropy. However, the preferred direction of the exciton
$E_\Lambda$ is rotated by about 30 degrees. This is not surprising
since the main contribution of the exciton $E_\Lambda$ is from
those states at the $\Lambda$ points, which are not high-symmetry
points. Finally, because the prominent absorption peak at 1.4 eV
is the combination of these two excitons, it is obvious that the
overall optical absorption does not show a complete anisotropy as
shown in Fig.~\ref{fig:5}(d) with a 50-meV smearing. It has to be
pointed out that PL experiments may exhibit stronger anisotropic
optical spectra than the absorption measurements. At room
temperature, the 96-meV gap between the excitons $E_\Gamma$ and
$E_\Lambda$ will make most excited carriers collected by the
lower-energy $E_\Gamma$ exciton, resulting in a simpler
anisotropic PL intensity similar to the exciton shown in
Fig.~\ref{fig:5}(b).

\section{Summary}

In summary, first-principles GW-BSE simulations have been
performed to study quasiparticle band gaps, excitonic effects, and
optical spectra of distorted 1-T diamond-chain monolayer ReS$_2$ and ReSe$_2$.
Monolayer ReS$_2$ is direct band gap semiconductor, and the
quasiparticle band gap is 2.7 eV. Monolayer ReSe$_2$ changes from
the DFT-calculated 1.22-eV indirect band gap to be a 2.3-eV direct
one after GW corrections. Huge excitonic effects dominate optical
spectra of suspended monolayer ReX$_2$, with around 1-eV
\emph{e-h} binding energies. The calculated absorption peaks are
in excellent agreement with recent measurements. Additionally,
both the monolayer ReS$_2$ and ReSe$_2$ shows anisotropic optical
responses due to their reduced triclinic symmetry lattices. In
particular, the partially anisotropic optical absorption spectra
of monolayer ReSe$_2$ are attributed to the two adjacent excitons,
resulting from two nearby maxima at the valence band edge. Our
calculations provide necessary information to clarify experimental
measurements and our predictions are useful for broad
near-infrared applications of distorted 1-T phase 2D semiconductors.

\begin{acknowledgments}
We thank Vy Tran, Ruixiang Fei, and Han Wang for fruitful
discussions. This work is supported by National Science Foundation
Grant No. DMR-1207141. H.-X. Z. and J.-J. S. are supported by the
National Basic Research Program of China (No. 2012CB619304) and
the National Natural Science Foundation of China (No. 11474012).
H.-X. Z. also acknowledges the financial support from the China
Scholarship Council. The computational resources have been
provided by the Lonestar of Teragrid at the Texas Advanced
Computing Center (TACC).
\end{acknowledgments}


\newpage

\begin{table}
\caption{\label{table1} DFT and GW band gaps (in unit of eV) for
monolayer ReS$_2$ and ReSe$_2$.}
 \begin{ruledtabular}
  \begin{tabular*}{6cm}{ccccc}
                 &DFT             & G$_0$W$_0$             & sc-G$_1$W$_0$             & sc-G$_1$W$_0$(SOC) \\
  \hline
  ReS$_2$        &1.36                &2.38              &2.82                        &2.69   \\
                 &1.43$^{\rm a}$      &                   &                             &       \\
  ReSe$_2$       &1.22                &2.09              &2.45                        &2.29     \\
                 &1.24$^{\rm b}$      &                   &                             &    \\
\end{tabular*}
\end{ruledtabular}
\footnotetext[1]{Reference~\onlinecite{tongay2014monolayer}.}
\footnotetext[2]{Reference~\onlinecite{yang2014layer}.}
\end{table}

\begin{table}
\caption{\label{table2} The PDOS of the highest valence and lowest
conduction band wave functions for monolayer ReS$_2$ and ReSe$_2$
at the $k$-points labelled in Fig.~\ref{fig:2}. The percent
contributions from each atomic orbital for these wave functions
are listed.}
 \begin{ruledtabular}
  \begin{tabular*}{7cm}{cccccccccccc}
           &        &\multicolumn{2}{c}{Energy (eV)}                       & \multicolumn{3}{c}{Re}      & \multicolumn{2}{c}{X} \\
           &        &DFT  &G$_0$W$_0$            &$s$   &$p$      &$d$              &$s$    &$p$ \\
  \hline
  ReS$_2$  &$\Gamma_v$    &0.00     &0.00             &0     &5        &80       &3      &11              \\
           &$\Gamma_c$    &1.36     &2.38            &1     &2       &69        &1      &27               \\
  ReSe$_2$ &$\Gamma_v$    &-0.02    &0.02            &0     &6        &78       &0      &16            \\
           &$\Gamma_c$    &1.22     &2.11            &0     &1        &59      &1      &39              \\
           &$\Lambda_v$   &0.00     &0.00            &0     &4        &86       &0      &10             \\
           &$\Lambda_c$   &1.29     &2.18             &0     &1        &63      &1      &35         \\
\end{tabular*}
\end{ruledtabular}
\end{table}

\begin{figure}
\includegraphics[scale=0.5]{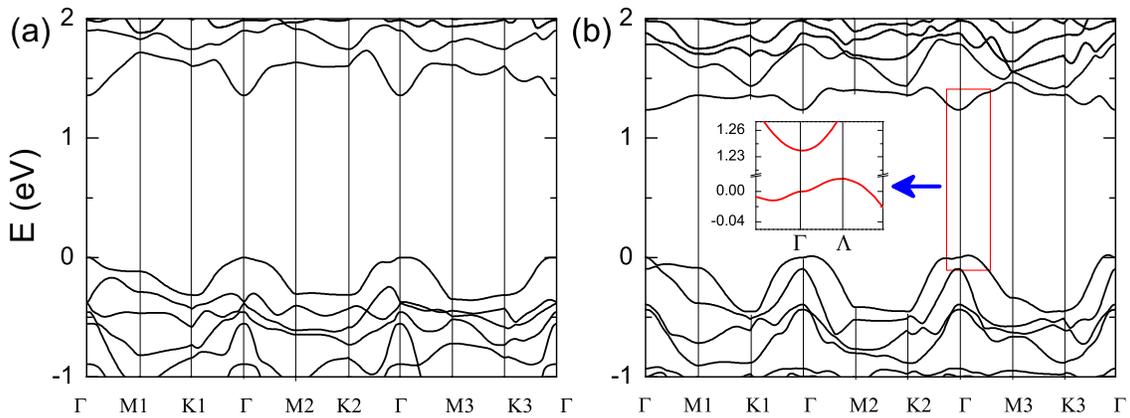}
\caption{The DFT-calculated band structures of monolayer ReS$_2$
(a), and ReSe$_2$ (b). The inset in (b) is the zoom of the band
structure inside the red rectangle. The SOC is not included in
these plots. The top of the valence band is set to be zero.}
\label{fig:2}
\end{figure}

\clearpage
\appendix*
\section{Band structure of monolayer ReX$_2$ with SOC included}
\begin{figure}
\includegraphics[width=10.0cm]{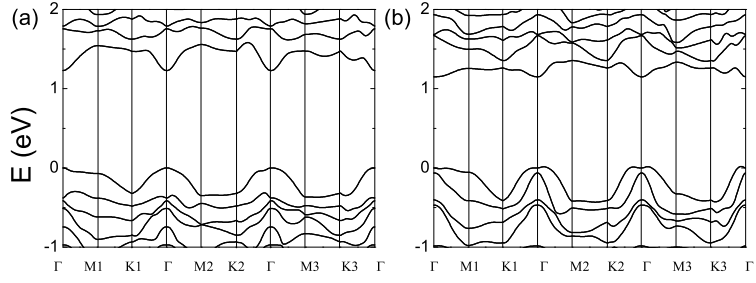}
\caption{The DFT-calculated band structures of monolayer ReS$_2$
(a), and ReSe$_2$ (b) with SOC included. The energy of the top of
the valence band is set to be zero.}\label{fig:6}
\end{figure}

Since Re atoms are heavy atomic species, we thus take into
accounts of SOC in Fig.~\ref{fig:6}. Compared with the band
structures in Fig.~\ref{fig:2}. the SOC slightly reduces the band
gap of monolayer ReX$_2$, which is also clearly presented in Table
I. \clearpage

\end{document}